\documentstyle[11pt,newpasp,twoside,epsf]{article}

\def\edcomment#1{\iffalse\marginpar{\raggedright\sl#1\/}\else\relax\fi}

\begin{document}
\title{The Chemical Evolution of Galaxy Disks}
\author{Francesca Matteucci}
\affil{Department of Astronomy, University of Trieste,
Via G.B. Tiepolo 11, 34100 Trieste, Italy}

\begin{abstract}
We discuss the main ingredients necessary to build models of chemical
evolution of spiral galaxies and in particular the Milky Way galaxy.
These ingredients include: the star formation rate, the initial mass
function, the stellar yields and the gas flows.  Then we discuss
models for the chemical evolution of galaxy disks and compare their
predictions with the main observational constraints available for the
Milky Way and other spirals.  We conclude that it is very likely that
the disk of our Galaxy and other spirals formed through an
``inside-out'' mechanism, where the central parts collapsed much
faster than the external ones. This mechanism has important
consequences for the appearance of galaxy disks as a function of
redshift.
\end{abstract}

\section{Introduction}
In order to predict the chemical evolution of galaxy disks one needs
to know several fundamental quantities. In particular, to build a
model which predicts the evolution of the abundances of the most
common chemical elements in the gas as functions of space and time,
some assumptions on the following physical processes are required: i)
the star formation, ii) the stellar nucleosynthesis, iii) the possible
gas flows entering and/or leaving the system. Unfortunately, point i)
is still very uncertain since we do not know the physics of star
formation and we have to adopt simplifying assumptions about the star
formation law.  Generally we divide the stellar birthrate into two
independent functions, one is the so-called star formation rate (SFR)
and is only a function of time, the other is the initial mass function
(IMF) and is only a function of the stellar mass. Concerning point
ii), the knowledge of stellar evolution and nucleosynthesis is more
advanced than that of the star formation process although large
uncertainties still exist also in the calculation of the stellar
yields; among these there are the treatment of convection, the rates
of nuclear reactions and the explosive nucleosythesis. In particular,
the yield of iron from massive stars ($M > 10 M_{\odot}$) seems to be
particularly uncertain owing to the poorly known ``mass-cut'', namely
how much mass stays in the collapsed remnant and how much is ejected
in the supernova event. Finally, point iii) is also highly uncertain
since the gas dynamics is a very complex phenomenon and pure chemical
evolution models take into account dynamics only in a very simplified
manner.

In past years a great deal of theoretical work has appeared concerning
the chemical evolution of the disk of the Milky Way and other spirals
(Matteucci and Fran\c cois 1989; Yoshii and Sommer-Larsen, 1990;
Burkert et al. 1992; Carigi 1994; Timmes et al. 1995; Tsujimoto et
al. 1995; Chiappini et al. 1997, 1999,2000a,b; Chang et al. 1999;
Portinari \& Chiosi 1999, 2000, Boissier \& Prantzos 1999; Goswami \&
Prantzos 2000;Prantzos and Boissier, 2000; Romano et al. 2000).  Most
of the more recent models agree that is important to assume an
``inside-out'' formation for disks. This implies that the inner parts
of galaxy disks should form on relatively fast timescales whereas the
outermost regions might be still forming now.

In this paper we focuse on one particular model for the chemical
evolution of the Galactic disk, the model of Chiappini et al. (1997,
hereafter CMG97) and its more recent development (Chiappini et
al. 2000b).  This model assumes two main episodes of gas infall of
primordial chemical composition. The first episode occurs on a short
timescale ($\sim 0.8 $ Gyr) and contributes to the formation of the
halo, the second takes much longer (several Gyrs) and gives rise to
the disk.  In this picture the formation of most of the disk is
disentangled from that of the halo which is assumed to have lost part
of its gas to form the bulge.  Therefore, the bulge formed on a
timescale similar to that of the halo.  The formation of the innermost
disk regions also occurs on a short timescale whereas the outermost
regions form on much longer timescales.  >From the observational point
of view, the distributions of abundances, stars and gas in the disk of
our Galaxy are the best studied.  In particular, the presence of
abundance gradients along the Galactic disk is now well established.
Abundance gradients represent an important constraint on the history
of the formation of the disk especially if considered together with
the distribution of gas, stars and star formation as functions of the
Galactocentric distance.  Abundance gradients are also measured in
other spirals and in some cases the gas distribution is also
available.  Finally, in the local disk (solar neighbourhood) the
abundances in stars of different ages are very well studied.  Very
good constraints are represented by the G-dwarf metallicity
distribution which imposes constraints on the birthrate history of the
local disk and the [el/Fe] versus [Fe/H] relations which help us in
understanding the stellar nucleosynthesis and supernova (SN)
progenitors.

\section{Observational constraints for the Milky Way Disk}

\subsection{The Local Disk}
\begin{itemize}
\item The G-dwarf metallicity distribution (Wyse and Gilmore 1995;
Rocha-Pinto and Maciel, 1996), represents the history of the stellar
birthrate in the local disk and it can impose strong constraints on
the mechanisms of disk formation.  In particular, the fit of the
G-dwarf [Fe/H] distribution requires that the local disk formed by
infall of gas on a time scale of the order of $\tau_d \sim 6-8$ Gyr,
(CMG97; Romano et al. 2000; Boissier and Prantzos 1999).

\item The relative abundance ratios as functions of the relative
metallicity (relative to the Sun) [X/Fe] vs. [Fe/H] are generally
interpreted to be due to the time-delay between Type Ia and II SNe.
From these abundance patterns one can infer the timescale for the
halo-thick disk formation ($\tau_h \sim$ 1.5-2.0 Gyr, Matteucci and
Fran\c cois, 1989; CMG97).

\item Evidence for a hiatus (no longer than 1 Gyr) in the SFR before
the formation of the thin disk is suggested by the increase of the
[Fe/$\alpha$] (with $\alpha$=Mg and O) ratio at a constant
[$\alpha$/H] value, (Gratton et al., 2000; Furhmann, 1998).

\item The age-metallicity relation (Edvardsson et al. 1993). This
relation is not a strong constraint since it can be fitted by a
variety of model assumptions and it shows a very large spread. Part of
this spread can be due to the fact that in the local disk we observe
also stars which are born elsewhere in the disk and later migrated
into the solar vicinity because of orbital diffusion (e.g Wielen et
al. 1996).
\end{itemize}

\subsection{The Whole Disk}

The main observational constraints for the whole disk are:
\begin{itemize}
\item Abundance Gradients: data from various sources (HII regions,
PNe, B stars) suggest a standard value for the gradient for oxygen,
$\sim -0.07$ dex/kpc in the galoctocentric distance range 4-14 kpc
(Maciel and Quireza 1999 and references therein).  However, it is not
yet clear if the slope is unique or bimodal. Similar gradients are
found for N and Fe.

\item Gas Distribution :HI is roughly constant over a range of 4-10
kpc whereas $H_2$ follows the light.  The total gas increases towards
the center with a peak at 4-6 kpc.

\item SFR Distribution: from various tracers (Lyman-$\alpha$
continuum, pulsars, SN remnants, molecular clouds) one finds that the
SFR increases towards the inner disk regions with a peak at 4-6 kpc
similar to that of the gas.

\end{itemize}

It is worth noting that most of the galactic chemical evolution models
suggest, in agreement with Larson (1976), that in order to fit
abundance gradients, SFR and gas distribution along the Galactic disk,
an inside-out formation of the disk is required, with the timescale
for the formation of the disk being a linearly increasing function of
the Galactocentric distance, and that the SFR should be a strongly
varying function of the Galactocentric distance.

\section{Observational constraints on disks of other spirals}

\begin{itemize}

\item Abundance gradients: if measured in dex/kpc are steeper in
smaller disks but the correlation disappears if dex/$R_d$ is used,
namely there is a universal slope per unit scale length (Garnett,
1998).  Another important characteristic is that the gradients seem to
be flatter in galaxies with central bars (Zaritsky et al. 1994).

\item The SFR: is measured mainly from $H_{\alpha}$ emission
(Kennicutt, 1989; 1998) and shows a well defined correlation with the
total surface gas density (HI+$H_{2}$).

\item Gas distribution: there are differences between field and
cluster spirals indicating that these latter have suffered
environmental effects (Skillman et al., 1996).

\item Integrated colors show that a formation inside-out is required
(Jimenez et al. 1998; Prantzos and Boissier 2000), in agreement with
what is inferred for the Milky Way disk.
\end{itemize}

\section{Different approaches to the chemical evolution of the Galactic disk}
The most common approaches to the formation of the Galactic disk and
disks in general, from a chemical evolution point of view, are:
\begin{itemize}

\item Serial Formation: in this scenario halo, thick and thin disk
form in sequence (e.g. Matteucci \& Fran\c cois 1989; Burkert et
al. 1992).  No overlapping in metallicities between the different
stellar populations, at variance with observational evidence (Beers
and Sommar-Larsen 1995) is predicted.

\item Parallel Formation: the various Galactic components start at the
same time and from the same gas but evolve at different rates. Within
this scenario one predicts overlapping of stars belonging to the
different components (e.g. Pardi et al. 1995). However, a problem of
this approach, common also to the previous one, is that it is
difficult to disentangle the evolution of the halo from that of the
disk, at variance with suggestions from the distribution of the
angular momentum of stars in the different components (Wyse and
Gilmore, 1992), indicating that the gas which formed the stellar halo
did not participate in the formation of the disk.

\item Two-infall Formation: the evolution of the halo and thin disk
are independent and they form out of two separate infall episodes
(overlapping in metallicity of different stellar populations is also
predicted) (e.g. CMG97; Chang et al. 1999).
\end{itemize}

\section{Basic Ingredients for Galaxy Evolution}

\subsection{The Star Formation Rate}
Given the poor knowledge of the physical processes determining star
formation we are forced to parametrize the SFR. Various
parametrizations have been proposed, in particular:

\begin{itemize}

\item Exponentially decreasing with time: $SFR= \nu e^{-t/ \tau_*}$,
with $\tau_* = 5-15$ Gyr (Tosi, 1988) in order to give good agreement
with the present time gas and SFR in the solar neighbourhood.

\item The Schmidt law, namely depending on a power between 1 and 2 of
the volume or surface gas density: $SFR=\nu \rho^{k}_{gas}$ or $SFR =
\nu \sigma_{gas}^{k}$.

\item A law which includes the total surface mass density: $SFR= \nu
\sigma_{tot}^{k_1} \sigma_{gas}^{k_2}$, where $\sigma_{tot}$ is the
total surface mass density and accounts for the feedback mechanism
between star formation and heating of the interstellar medium (ISM),
due to supernovae and stellar winds.  Observational evidence for such
a law is provided by Dopita and Ryder (1994).

\item A function of the surface gas density and the angular rotation
speed of the gas: $SFR= 0.017 \Omega_{gas} \sigma_{gas}\propto R^{-1}
\sigma_{gas}$, with $\Omega_{gas}$ being the angular rotation speed of
the gas. This law was suggested by Kennicutt (1998) as a good fit of
the SFR measured from the $H_{\alpha}$ emission in spirals and
starburst galaxies.  Kennicutt has also suggested the existence of a
threshold gas density for star formation of few $M_{\odot}pc^{-2}$,
below which the star formation cannot occur. CMG97 have adopted such a
thrshold in their model and have shown that this naturally produces a
hiatus in the star formation rate between the thick and thin disk
formation as observed in the plots [Fe/Mg] vs. [Mg/H] and [Fe/O]
vs. [O/H] (Fuhrmann 1998; Gratton et al. 2000).

\end{itemize}

All of these SFRs have to be calibrated in order to reproduce the
present time SFR in the solar vicinity and this is done by means of
the parameter $\nu$ which is the efficiency of star formation and is a
free parameter.

The local disk SFR, measured from the tracers mentioned above and by
assuming a specific IMF, is in the range SFR=2-10$M_{\odot}pc^{-2}
Gyr^{-1}$ (Timmes et al. 1995).

\subsection{The Initial Mass Function}

The IMF is normally parametrized as a power law of the type:
$\varphi(m) \propto m^{-(x+1)}$, defined over a mass range of
0.1-100$M_{\odot}$.  There is a general agreement that the Salpeter
IMF with $x=1.35$ does not work well for the disk of the Galaxy and
that IMFs with steeper slopes in the range of massive stars have to be
preferred (e.g. Scalo, 1986).  On the other hand, it seems that in
elliptical galaxies a Salpeter or even flatter IMF gives better
results elaborating the evolution of these galaxies and the
intra-cluster medium (see Matteucci 1996).

\subsection{The Infall Rate}

The importance of the gas infall in the formation of the Galactic disk
is originally due to the best fit that it provides for the G-dwarf
metallicity distribution in the solar neighbourhood. A biased gas
infall (inside-out formation) also provides a very good explanation
for the main properties of the disk (abundance gradients, gas and star
formation) as functions of the Galactocentric distance.

There are various parametrizations of the infall rate:

\begin{itemize}
\item Constant in space and time.

\item Variable in space and time: $IR= A(R) e^{-t/ \tau(R)}$, with
$\tau(R)$ constant or varying along the disk.  The quantity $A(R)$ is
derived by fitting the actual surface mass density distribution along
the disk: $\sigma_{tot}(R) = \sigma_o e^{-R/R_d}$.

\item CMG97 adopted a double law for the halo-thick disk and
thin-disk: $IR= A(R) e^{-t/ \tau_{H}} + B(R)
e^{-(t-t_{max})/\tau_{D}(R)}$, where $\tau_H$ and $\tau_{D}(R)$ are
the timescales for the halo-thick and thin-disk formation,
respectively.  The dependence of $\tau_{D}$ on the Galactocentric
distance is a linear function (see CMG97).
\end{itemize}

\subsection{Radial Flows}
Observationally, it is not clear whether radial flows along the
Galactic disk and disks in general really exist.  In a recent study,
Portinari and Chiosi (2000) analysed in detail the effects of radial
flows and concluded that they are important in connection with the
peak observed in the gas in the Galactic disk at 4-6 kpc.

\subsection{Stellar Nucleosynthesis}

We briefly recall here the element production by stars of all masses,
as suggested by the most recent nucleosynthesis calculations.  In
particular these calculations are: Woosley and Weaver (1995) and
Thielemann et al. (1996) for massive stars, Marigo et al. (1996) and
van den Hoeck and Groenewegen (1997) for low and intermediate mass
stars.

\begin{itemize}

\item Massive stars: $M >10 M_{\odot}$: produce the bulk of O and
$\alpha$-elements (Mg, Si, Ca, Ti) and some Fe (very uncertain). They
die as Type II SNe.

\item Low and Intermediate Mass Stars ($0.8 \le M/M_{\odot} \le 6-8$):
i) single stars produce C, N and some s-process elements, they die as
C-O white dwarfs (WDs).  ii) Carbon-oxygen white dwarfs in binary
systems give origin to Type Ia SNe when they accrete enough matter
from the companion to reach the Chandrasekhar mass limit.  At this
point a C-deflagration is initiated with a consequent explosion (Type
Ia SN), and the WD is transformed into Fe and traces of elements from
C to Si. They are responsible for the production of the bulk of Fe in
galaxies, unless a very unusual IMF is adopted.

\end{itemize}

\section{Model Results}

As an example of a model which can satisfactorily fit the properties
of the Galactic disk we show the results of Chiappini et al. (2000b)
and Chiappini et al. (this conference).  The model is basically that
of CMG97 except that here we considered a more realistic density
profile decreasing outwards for the stellar halo rather than constant
as in CMG97.  The nucleosynthesis prescriptions are the most recent
ones and the nucleosynthesis of novae is also included.  In figure 1
we show the predicted oxygen abundance gradient at the present time,
the distribution of the $SFR$/$SFR_{\odot}$ ratio, the total surface
gas density and the stellar density distributions.

\begin{figure}
\plotone{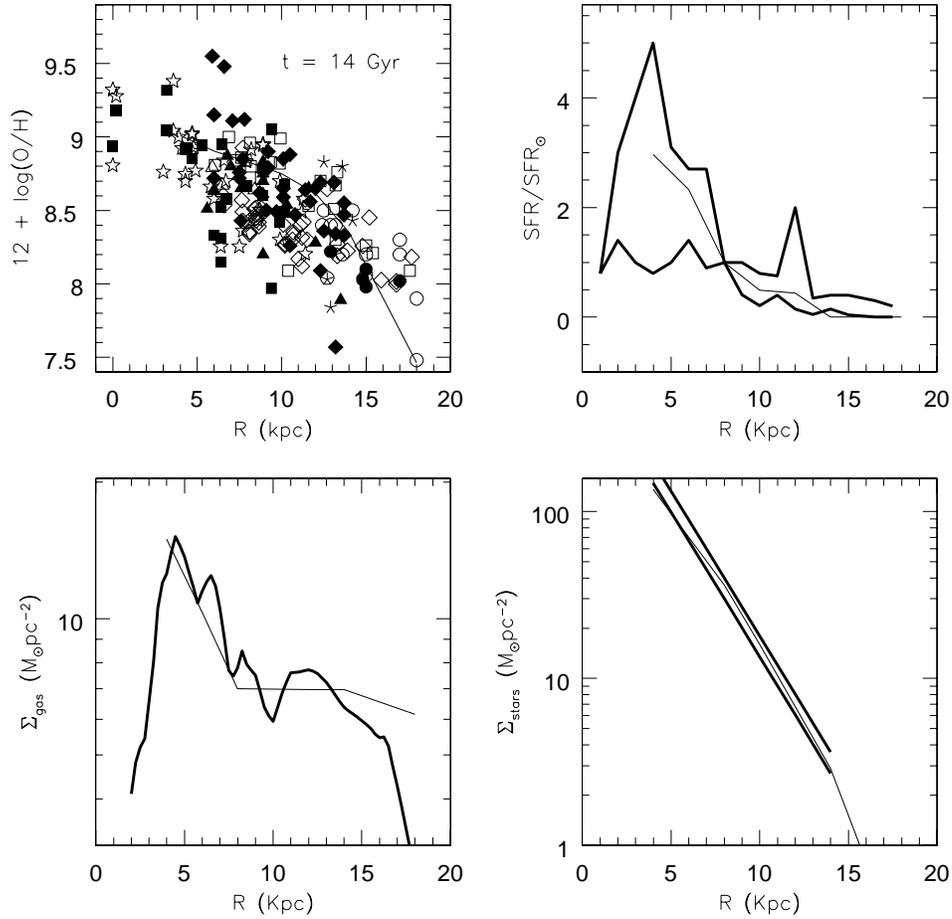}
\caption{Predictions (thin lines) from the best model of Chiappini et
al. (2000b).  In the upper left panel is shown the oxygen gradient
compared to the data, in the upper right panel the distribution of the
SFR, in the lower left panel the distribution of the total surface gas
density and in the lower right panel the distribution of stars.  The
thick lines enclose the areas where the observations lie (see
Chiappini et al., 2000b for references).}
\vspace {2.5in}
\end{figure}

The model reproduces all the four constraints except for the peak of
the gas and SFR at 4-6 kpc which is probably the result of the
dynamical effects of the central bar, as discussed in Portinari and
Chiosi (2000).

\section{Conclusions}

The comparison between observations and models suggests that:

\begin{itemize}

\item The disk of the Galaxy formed mostly by infall of primordial or
very metal poor gas accumulating faster in the inner than in the outer
regions (inside-out scenario).

\item In the framework of the inside-out scenario the SFR should be a
strongly varying function of the galactocentric distance.  This can be
achieved by assuming a dependence of the SFR either on the total
surface mass density (feed-back mechanism) or on the angular circular
velocity of gas (both are supported by observations), besides the
dependence upon the surface gas density.

\item Radial flows probably are not the main cause of gradients but
can help in reproducing the gas profile.

\item An IMF constant in space and time should be preferred, as shown
by several numerical experiments (e.g. Chiappini et al. 2000a).

\item The disks of other spirals also indicate an inside-out
formation, as shown by Prantzos and Boissier (2000) who described the
galaxy disks by means of scaling laws, taken from semy-analytical
models of galaxy formation (Mo et al. 1998), calibrated on the Galaxy.

\item Dynamical processes such as the formation of a central bar can
influence the evolution of disks and deserve more attention in the
future.

\item Abundance ratios (e.g. [$\alpha$/Fe]) in stars at large
galactocentric distances can provide clues for understanding the
formation of the disk and the halo (inside-out or outside-in) (see
Chiappini et al., 2000b).

\item The predictions of chemical evolution models can be tested in a
cosmological context by studying galaxy surface brightness and size
evolution as a function of redshift. Roche et al. (1998) concluded
that a size-luminosity evolution, as predicted by the inside-out model
of CMG97, represents a good fit to the observations.
\end{itemize}

\end{document}